\documentclass[twocolumn, times]{aastex63}

\usepackage{amsmath,verbatim, nicefrac,color, soul}
\usepackage{gensymb}
\usepackage{natbib,graphicx}
\usepackage{array}

\usepackage[caption=false]{subfig}

\makeatletter
\newcommand\notsotiny{\@setfontsize\notsotiny\@vipt\@viipt}
\makeatother

\newcommand\Hl[1]{\colorbox{yellow}}

\DeclareCaptionFormat{cont}{#1 (cont.)#2#3\par}

\shorttitle{Scattering Medium Geometry with Cyclic Spectroscopy}
\shortauthors{J. E. Turner}

\begin{document}

\title{Constraining Scattering Medium Geometry with Cyclic Spectroscopy} 

\author[0000-0002-2451-7288]{Jacob E. Turner}
\affiliation{Green Bank Observatory, P.O. Box 2, Green Bank, WV 24944, USA}

\begin{abstract}
We use cyclic spectroscopy to infer the scintillation parameter $C_1$ for the millisecond pulsar B1937+21. This marks the first time this parameter has been measured for any pulsar without assuming a pulse broadening function shape prior to deconvolution from the intrinsic pulse profile, removing significant potential biases in scattering delay estimation and letting us consider a wider range of line-of-sight geometries. At 428 MHz, we find a weighted mean and standard deviation across multiple epochs and independent scintillation fluctuations of $C_1$=1.21$\pm0.01$, which, along with the consistent presence of diffuse scintillation arcs, favors a thick screen geometry spanning just over 10\% of the Earth-pulsar distance. The resulting precision in our weighted average allows us to rule out various thin screen geometries, as well as thick screen geometries comprising more than 30\% of the Earth-pulsar distance, with greater than $5\sigma$ certainty at this observing frequency. We also use our measured $C_1$ values to determine diffraction scales, which we find to be roughly 11$\times10^3$ km between 418 and 438 MHz, suggesting an inner scale on the order of $10^3$ km. Future implementations of our method to other lines of sight through the galaxy may substantially improve efforts to understand structures that contribute to the majority of pulsar emission scattering in the interstellar medium. As flagship instruments like the Green Bank Telescope begin offering the use of cyclic spectroscopy backends, and other instruments begin exploration and commissioning of similar systems, demonstrations like these will be crucial for the widespread adoption of cyclic spectroscopy. 
\end{abstract}

\section{Introduction}\label{intro}
Cyclic spectroscopy \citep{cyc_spec} is a signal processing technique that takes advantage of periodicity in signals, such as those naturally occurring in pulsar emission viewed from Earth, to extract additional information via their harmonics that would be otherwise difficult to attain through typical phase-resolved spectroscopy. For a pulsar signal, $s(t)$, with Fourier transform, $S(\nu)$, that passes through a linear filter such as the ionized interstellar medium (ISM), the signal that reaches Earth will be a convolution of the intrinsic pulsar signal with the scattering impulse response of the ISM, $h(t)$, known as the pulse broadening function. The corresponding cyclic spectrum is given by 

\begin{equation}
    S_{E}(\nu,\alpha_k) = \langle H(\nu+\alpha_k/2)H^*(\nu-\alpha_k/2)\rangle S(\nu, \alpha_k),
\label{pulse_cs}
\end{equation}

\noindent where $S_{E}$ denotes the cyclic spectrum for the total electric field signal observed at Earth, $\alpha = k/P$ where $k$ is an integer and $P$ is the pulse period, $H(\nu)$ is the transfer function, also known as the frequency response, of the ISM, and the expectation value (angle brackets) is considered over time intervals before the signal decorrelates due to scattering, known as the diffractive timescale or scintillation timescale. 

\par In pulsar data, the transfer function contains a diffraction pattern that forms due to pulsar emission undergoing a multipath propagation through an ionized medium prior to reaching Earth, and and the squared modulus of this function over both observing frequency and time is an observation's dynamic spectrum. The characteristic width in observing frequency of fringes in this diffraction pattern can be described via $\Delta \nu_{\rm d}$, known as the diffractive bandwidth or scintillation bandwidth \citep{Cordes_1998}. Similarly, we can characterize this behavior in the time domain via the first moment of the pulse broadening function, $\tau_{\rm d}$, which we will refer to throughout the remainder of this paper as the scattering delay, also originating from the multipath nature of pulsar emission. In the case of medium with a square-law structure function, $\tau_{\rm d}$ can be fully expressed as the sum of delays, $\tau_{n}$, which are solutions to eigenvalue equations associated with the convolution of $n$ broadening functions describing the entire line of sight propagation,

\begin{equation}
   \tau_{\rm d} = \sum_n \tau_{n}.
\end{equation}

\noindent This expression can be reduced to $\tau_{\rm d} \approx \tau_{1}$ in instances where a singular, localized region of the line of sight dominates the scattering, such as the case of a thin screen, which can be characterized by a one-sided exponential pulse broadening function \citep{Cordes_1998}. The shape of $h(t)$ itself is dependent on the form of its constituent broadening functions, and can change significantly based on the electron density wavenumber spectrum and how the ISM structure that induces scattering is distributed between observer and source, i.e., the line of sight geometry \citep{geiger2024nanograv125yeardataset}. 

\par If one can simultaneously measure $\Delta \nu_{\rm d}$ and $\tau_{\rm d}$, it is possible to infer scattering medium geometry via the scintillation parameter, $C_1$, given by
\begin{equation}
\label{c1_eq}
C_1 = 2\pi \Delta \nu_{\text{d}} \tau_{\rm d}.
\end{equation}
Consequently, constraining $C_1$ can be crucial for narrowing down potential features in the ISM from which scattering originates, which itself remains an open question. While some pulsar lines of sight have scattering that can be traced back to well-determined, localized structures, i.e., thin screens, such as HII regions \citep{mall}, the boundary of the Local Bubble \citep{McKee_2022}, bow shock nebulae \citep{reardon_nature}, or hydrogen filaments \citep{stock}, many pulsars, even those fairly local to Earth (within 1 kpc), have localized scattering screens with no clear associated ISM feature \citep{ocker}. Interestingly, there also exist lines of sight through the galaxy where the scattering appears to be uniformly distributed \citep{Marthi_2025}. 
\par Additionally, $C_1$ provides a direct means through which one can relate diffractive bandwidth to scattering delay. These diffractive bandwidths are monitored by pulsar timing arrays (PTAs) as a source of timing noise that can lessen sensitivity in gravitational wave detection efforts \citep{Levin_Scat, turner_scat, epta_scint}, and are estimated to be similar to or greater than median pulse time-of-arrival uncertainties in a non-negligible fraction of sources \citep{Agazie_2023_timing,Turner_1937}. However, PTAs monitor scattering delays via diffractive bandwidth measurements, and need to assume a value for $C_1$ (typically 0.957 or 1, which are thin screen geometries assuming either a Kolmogorov or square-law medium, respectively) for conversion. Consequently, deviations between the assumed and true line of sight geometry can result in significant misestimations of scattering delay, biasing PTA noise budgets.  

\par While incredibly valuable as a tool for characterizing lines of sight through the galaxy, $C_1$ is difficult to recover in practice. First, scintillation bandwidth and scattering delay, usually on the order of kHz$-$MHz and ns$-\mu$s respectively, are typically only independently measurable in different frequency regimes due to resolution limitations described by the time-frequency sampling uncertainty relation. Second, the shape of $h(t)$ is generally not known a priori for any line of sight, necessitating an assumed shape, and therefore an assumed line of sight geometry and wavenumber spectrum, for any attempted deconvolution from the intrinsic pulse profile. This biases any resulting estimated $\tau_{\rm d}$, and can only accurately describe the line of sight geometry if the scattering highly localized, otherwise requiring a recursive deconvolution of many broadening functions, of which the appropriate number is, also, not known a priori \citep{Marthi_2025}. Finally, since these deconvolutions only utilize the intensity of the pulsar signal, and thus only its amplitude, any phase information present is ignored and, consequently, only the envelope of the associated broadening function is recovered, such as in the CLEAN algorithm \citep{Bhat_2003, Young_2024}. As a result, while $C_1$ has been directly measured or alluded to in a handful of studies \citep{backer_vela, c1_measurement_1980, Main_2017, Marthi_2025}, they all suffer from at least one of the aforementioned issues. 

\par Fortunately, techniques such as cyclic spectroscopy allow one to circumvent some of these problems. First, by resampling data modulo the pulse period, one can maintain high time (pulse phase) resolution while achieving frequency resolutions up to the pulsar spin frequency, allowing for simultaneous measurements of scintillation bandwidth and scattering delay \citep{wdv13}. Second, deconvolution via the cyclic spectrum is completely agnostic to the shape of $h(t)$, allowing for completely coherent descattering and bypassing any need for a potentially infinite recursive deconvolution of broadening functions \citep{cyc_spec}. This allows one to consider a wide array of possible line of sight geometries without biased results. Third, cyclic deconvolution recovers both the amplitude and phase of $h(t)$ rather than just the amplitude, meaning we can recover the entire pulse broadening function signal rather than just the envelope.

\par In this study, we use cyclic spectroscopy to perform the first ever inference of $C_1$ without the biasing assumptions associated with choosing a pulse broadening function shape prior to deconvolution, allowing for more accurate descriptions of line of sight geometry and turbulence through the galaxy. In Section \ref{data}, we describe our observing setup and data, and our justifications for their use in this effort. In Section \ref{analysis}, we detail our data reduction, processing, and acquisition of relevant observables. In Section \ref{results}, we discuss our results and the physical interpretations thereof. Finally, in Section \ref{conclusions}, we provide concluding remarks discussing the implications of our findings and suggestions for their application in future studies.

\section{Data}\label{data}
We used data from observing campaign P2067 (PI Demorest), described in \cite{cyc_spec} and subsequently used by \cite{wdv13} and \cite{Turner_2025}, consisting of raw voltage, dual polarization, baseband observations of the millisecond pulsar B1937+21 taken across three epochs (MJDs 53791, 53847, and 53873) with the ASP spectrometer on the Arecibo telescope. While observations on MJDs 53847 and 53873 consisted of a single observation with 4 MHz of bandwidth centered at 428 MHz, the observation on MJD 53791 consisted of data acquisition across six 4 MHz banks, with a total bandwidth spanning 418 MHz$-$442 MHz. Scans ranged in length from 40$-$140 minutes depending on the epoch and observing frequency.
\par Using these data assures that we are performing our analyses in this pulsar's full cyclic deconvolution regime \citep[$\sim$271$-$593 MHz;][]{Turner_2026}. Additionally, while the cyclic figure of merit \citep[a metric quantifying the likelihood of cyclic deconvolution;][]{dsj+20, Turner_2025} has not been calculated for this telescope-observing frequency combination, this source decisively passes the conservative cyclic merit threshold for the CHIME telescope, which is less sensitive, in the same frequency range, lending significant confidence that our analyses will yield complete cyclic deconvolution \citep{Turner_2026}.

\section{Analysis}\label{analysis}
\subsection{Data Processing}
All observations were processed using the cyclic spectroscopy option in \texttt{dspsr}\footnote{\url{https://dspsr.readthedocs.io/en/latest/}} \citep{dspsr}, creating data products with 4608 cyclic frequency channels, 1024 pulse phase bins, and 15 second subintegrations within individual 4 MHz polyphase filterbank channels. As with \cite{wdv13}, significant aliasing at the edges of filterbanks necessitated the removal of approximately 0.22 MHz in bandwidth from both edges of each polyphase filterbank channel. Additionally, on MJD 53791, significant falloff in S/N in the bottom half of the lowest frequency 4 MHz polyphase filterbank (centered at 420 MHz) and the upper half of the highest frequency 4 MHz polyphase filterbank (centered at 440 MHz) necessitated limiting our analyses to the upper and lower halves of their bandpasses, respectively. 
\par Cyclic spectra must be limited to data spanning at most a scintillation timescale, $\Delta t_{\rm d}$ \citep{wdv13}. While this limits S/N in recovered pulse broadening functions, since scintillation timescales are generally on the order of minutes for most pulsars, an observation spanning 1$-$2 hours may result in many tens of unique and independent $C_1$ measurements, which significantly improves our ability to constrain this quantity. We determined $\Delta t_{\rm d}$ for each epoch-observing frequency combination by creating dynamic spectra, taking 1D slices of each spectrum's autocorrelation function at zero time lag, and determining the half width at $1/e$ via a fit with a one-sided exponential. This resulted in measured scintillation timescales of roughly 79 seconds for MJD 53791, 74 seconds for MJD 53847, and 71 seconds for MJD 53873. From there, we divided our observations into subscans equivalent to those timescales rounded down to the closest whole multiple of subintegration. We chose this approach rather than reprocessing our data to subscans equivalent to our measured scintillation timescales out of an abundance of caution to guarantee each observation was divided into segments that did not exceed this quantity within uncertainty.

\begin{figure*}[htbp]
\centering
\includegraphics[width=0.495\textwidth]{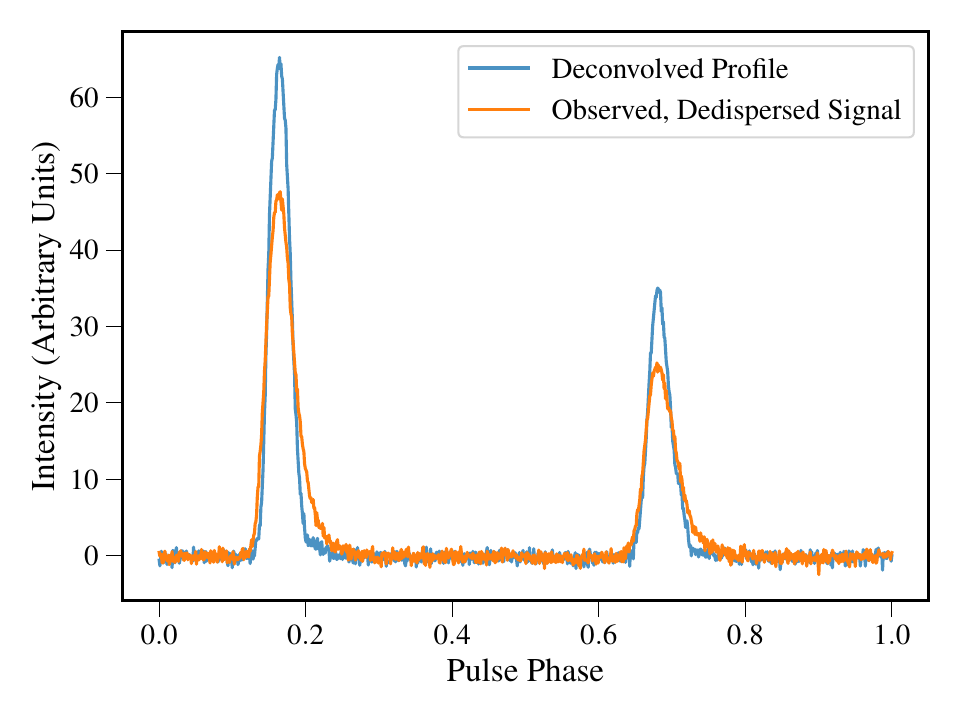}
\hfil
\includegraphics[width=0.495\textwidth]{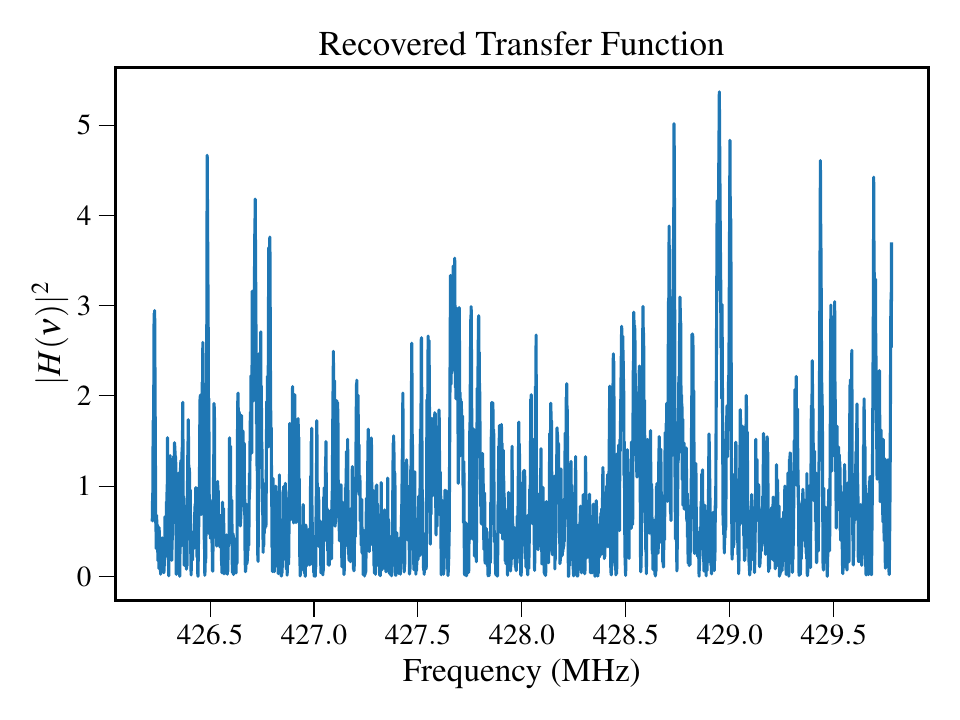}

\medskip
\includegraphics[width=0.495\textwidth]{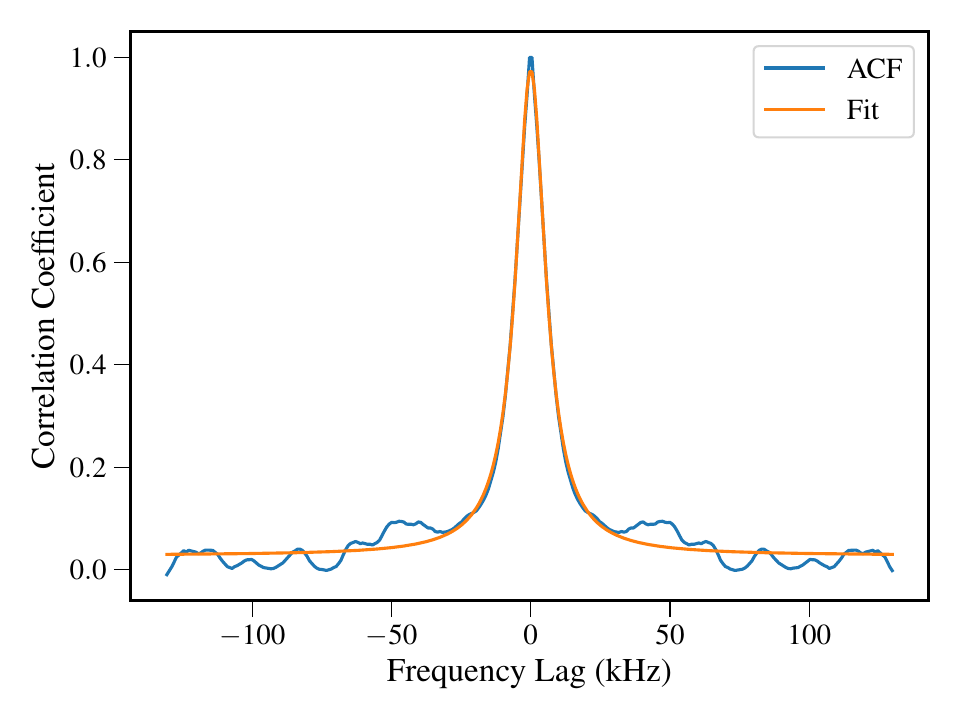}
\hfil
\includegraphics[width=0.495\textwidth]{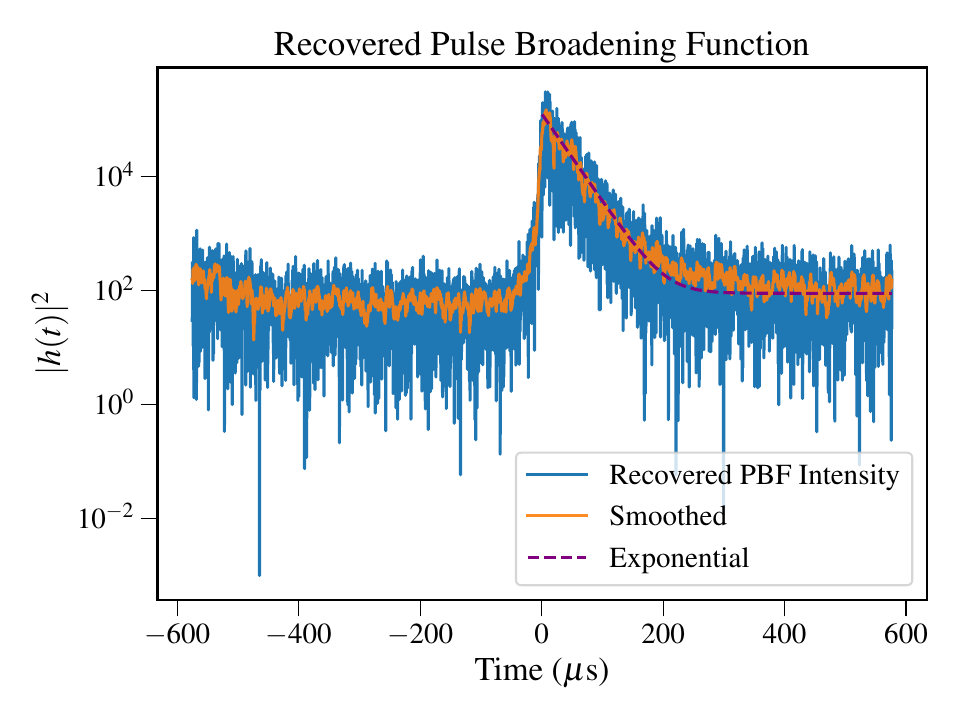}
 \caption{Recovered signals from a scintillation timescale's worth of total intensity data processed with cyclic deconvolution. (Top left) dedispersed pulse profile of PSR B1937+21 before (orange) and after (blue) deconvolution, each with 1024 phase bins. Note that, at higher observing frequencies, a narrow, trailing feature is present in both the main and interpulse components of this pulsar's profile, and its visibility at 428 MHz is significantly diminished due to profile evolution rather than insufficient pulse phase resolution (see \cite{wdv13}). (Top right) recovered transfer function intensity. (Bottem left) Autocorrelation function and fit for $|H(\nu)|^2$. (Bottom right) recovered pulse broadening function intensity (blue) with a time resolution of $\approx 0.28 \mu$s, the same function smoothed (orange), and a one-sided exponential with an equivalent scattering delay at a $1/e$ broadening time (dashed purple). Note the shorter tail on the exponential PBF compared to what was recovered for this observation, indicating disagreement with a thin screen interpretation.}%
    \label{pycyc_ex}%
\end{figure*}

\subsection{Cyclic Deconvolution \& Observable Determination}
\par Following data segmentation into subscans, each subscan was processed separately using the \texttt{pycyc}
\footnote{\url{https://github.com/gitj/pycyc}} python package, which, using the algorithm described in \cite{wdv13}, performs an iterative deconvolution via the cyclic spectrum to extract each subscan's transfer function/pulse broadening function. An example output from this process can be seen in Figure \ref{pycyc_ex}. Since these data had two polarizations, we performed deconvolution both with individual polarizations as well as total intensity. For each polarization configuration, we then measured each subscan's scintillation bandwidth by taking an autocorrelation function of the mean-subtracted transfer function intensity and determining its half width at half maximum via a Lorentzian fit \citep{Cordes_1998}. 

\par To measure a subscan's scattering delay, one could conceivably fit a function such as a one-sided exponential to the pulse broadening function. However, performing any such fit implicitly assumes a broadening function shape, which would bias our results and defeat the purpose of using a technique that is agnostic to its structure. To avoid such bias, for a broadening function spanning a range of delays, $t$, from 0 to $T$, we first centered the beginning of the broadening function at $t=0$ such that the signal ranged in time from $t=-T/2$ to $t=T/2$. From there, we determined the scattering delay using the weighted average delay of the pulse broadening function intensity. For a pulse broadening function, $h(t)$, this delay can be found via

\begin{equation}
\label{toa_cent}
\tau_{\textrm{d}} = \frac{\int_{-\frac{T}{2}}^\frac{T}{2} t \ |h(t)|^2dt}{\int_{-\frac{T}{2}}^{\frac{T}{2}} |h(t)|^2dt},
\end{equation}
where Equation \ref{toa_cent} would be equivalent to the $1/e$ decay time often used to analyze pulse broadening functions if $|h(t)|^2$ were a one-sided exponential. However, the broadening function shape and the length of the associated tail can differ dramatically depending on the wavenumber spectrum and scattering regime.

\par Uncertainty on the scintillation bandwidth consisted of the autocorrelation fit uncertainty and the finite scintle error added in quadrature, while the uncertainty on the scattering delay consisted solely of the finite scintle error. These finite scintle errors on scintillation bandwidth and scattering delay are defined as 
\begin{equation}
\label{finite_scintle}
\begin{split}
\epsilon_{\nu} &  = \frac{\Delta \nu_{\rm d}}{2\ln(2) N_{\rm scint}^{1/2}} \\
& \approx \frac{\Delta \nu_{\rm{d}}}{2\ln(2)[(1+\eta_{\text{t}}T/\Delta t_\text{d})(1+\eta_\nu B/\Delta \nu_\text{d})]^{1/2}}
\end{split}
\end{equation}

and 

\begin{equation}
\label{finite_scintle}
\begin{split}
\epsilon_{\tau} &  = \frac{\tau_{\rm d}}{N_{\rm scint}^{1/2}} \\
& \approx \frac{\tau_{\rm{d}}}{[(1+\eta_{\text{t}}T/\Delta t_\text{d})(1+\eta_\nu B/\Delta \nu_\text{d})]^{1/2}},
\end{split}
\end{equation}

\noindent respectively, where $N_{\rm scint}$ is the number of scintles present in the data, $T$ is the total observing length, $B$ is the observing bandwidth, and $\eta_{\rm t}$ and $\eta_{\nu}$ are filling factors generally ranging from 0.1 to 0.3, which we set to 0.2 \citep{Cordes1986}.

\par We note here that a recovered pulse broadening function may contain components originating from sources external to the ISM, such as impulse responses from instrumentation. However, given that any observation in which a tail can be clearly seen in the pulse profile exists firmly within the strong scattering regime, it is likely than any non-ISM components, if present, are statistically insignificant and can be ignored. Additionally, in some subscans, a small bump was occasionally present in the pulse broadening function beginning at half the total length of the delay range from the start of the primary recovered signal; this is due to the pulse profile of PSR B1937+21 containing an interpulse, which itself exhibits broadening. However, this signal is around 2$-$3 orders of magnitude smaller in intensity than the primary recovered signal when visible, and any attributable variation in measured scattering delay it produces between subscans, especially when considering that some inter-subscan variation will just be due to noise in our data, is smaller than the uncertainty given by the finite scintle effect. Additionally, this signal is distributed into positive and negative delays when the data are presented as spanning from $t=-T/2$ to $t=T/2$, further minimizing any potential bias. Finally, a perfectly recovered pulse broadening function would have a floor at zero intensity, particularly at non-physical, negative delays \citep{holography}, whereas our recovering pulse broadening functions have floors noticeably offset from zero because the observed signal will always contain some noise. However, the primary signal in our pulse broadening function intensities is generally 3$-$4 orders of magnitude above this floor, likely making any resulting biases negligible. We have confirmed this negligibility by examining the effects of offsets on scattering delays from recovered broadening functions, and found the resulting measured scattering delays to be minimally affected.

\par
\begin{figure}[!ht]
    \centering
    {\hspace*{-.5cm}\includegraphics[width=0.5\textwidth]{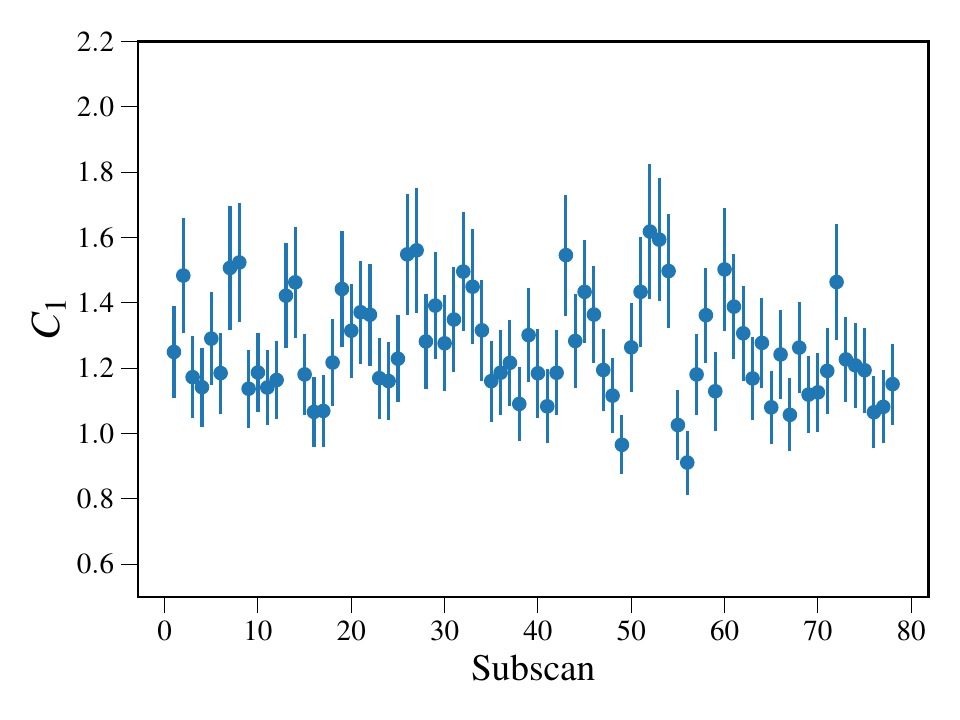}} %
    \caption{$C_1$ timeseries on MJD 53791 for the 4 MHz polyphase filterbank centered on 428 MHz.}
    \label{c1_timeseries}%
\end{figure}

 Once the scintillation bandwidth and scattering delay had been measured for a given data subscan, we calculated $C_1$ using Equation \ref{c1_eq}. This process was repeated for all subscans in a given epoch-observing frequency combination. An example timeseries showing recovered $C_1$ values from the 4 MHz polyphase filterbank centered at 428 MHz on MJD 53791 can be seen in Figure \ref{c1_timeseries}.

\break

\section{Results \& Discussion}\label{results}
\par The weighted means and weighted standard deviations of $C_1$, scintillation bandwidth, and scattering delay for all epoch-observing frequency combinations used in this study are shown in Table \ref{table_results}. 
We find significant agreement across our results, as all of our weighted average $C_1$ values agree to within 5\%, indicating strong stability in our measurements across both observing frequency and epoch.

\par \cite{Cordes_1998} utilize ray theory \citep{williamson_1975, blandford_1985} and wave theory \citep{lee_jokipii, ishimaru, Lambert_Rickett} to determine $C_1$ under various medium structure functions (square-law and Kolmogorov) and geometries (uniform, as well as screens and slabs of various thicknesses), the details and numerous examples of which are described in Section 3 and Table 2 of that study, respectively. We use these results as the basis of our $C_1$ conclusions throughout the remainder of this work. 

\par While some of our epoch-observing frequency combinations exhibit $C_1$ values that are consistent with a uniform geometry and a Kolmogorov medium \cite[$C_1$ = 1.16;][]{Cordes_1998}, the presence of consistent, diffuse scintillation arcs \citep{OG_arcs} in these observations \citep{wdv13,Turner_2025} indicates scattering is likely somewhat localized rather than completely uniformly distributed, potentially favoring a thick screen geometry.

\begin{deluxetable*}{CCCCCCC}
\tablecolumns{7}

\tablecaption{Measured \& Derived ISM Quantities \label{table_results}}
\tablehead{ \colhead{MJD} &
\colhead{Frequency} &  \colhead{Polarization} & \colhead{$\overline{C_1}$} & 
\colhead{$\overline{\Delta \nu_{\text{d}}}$} &
\colhead{$\overline{\tau_{\text{d}}}$} & \colhead{$N_{\text{meas}}$} 
 \\  
\colhead{} & \text{(MHz)} & \colhead{} &\colhead{} & \colhead{(kHz)}& \colhead{$(\mu\text{s})$} & \colhead{}} 
 
\startdata
53791 & 420.89 & \textrm{Total Intensity} &  1.22 $\pm$0.04 & 6.4$\pm$0.2 & 30.2 $\pm$0.5 & 35\\
53791 & 424 & \textrm{Total Intensity} & 1.25 $\pm$ 0.03 & 6.3$\pm$ 0.2 & 31.0$\pm$0.4 & 29\\
53791 & 428 & \textrm{Total Intensity} & 1.21 $\pm$ 0.02 & 6.2 $\pm$ 0.1 & 31.1$\pm$0.2 & 78\\
53791 & 432 &\textrm{Total Intensity} & 1.21 $\pm$0.03 & 6.5 $\pm$0.1 & 29.3 $\pm$0.3 & 51\\
53791 & 436 &  \textrm{Total Intensity} & 1.21 $\pm$0.03 & 7.2$\pm$0.2 & 26.4$\pm$0.4 & 35\\
53791 & 439.11 & \textrm{Total Intensity} & 1.15$\pm$0.03 & 5.9$\pm$0.2 & 30.4$\pm$0.4 & 55\\
53847 & 428 & \textrm{Total Intensity} & 1.22$\pm$0.02 & 7.9$\pm$0.1 & 24.4$\pm$0.2 & 76\\
53873 & 428 & \textrm{Total Intensity} & 1.19$\pm$0.02 & 7.2$\pm$0.1 & 25.7$\pm$0.2 & 117\\
53791 & 420.89 & \textrm{Polarization 1} &  1.19 $\pm$0.04 & 6.3$\pm$0.2 & 29.5 $\pm$0.5 & 35\\
53791 & 424 &  \textrm{Polarization 1} & 1.20 $\pm$ 0.03 & 6.3$\pm$ 0.2 & 30.2$\pm$0.4 & 29\\
53791 & 428 &  \textrm{Polarization 1} & 1.17 $\pm$ 0.02 & 6.1 $\pm$ 0.1 & 30.4$\pm$0.2 & 78\\
53791 & 432 &  \textrm{Polarization 1} & 1.17 $\pm$0.03 & 6.4 $\pm$0.2 & 28.9 $\pm$0.2 & 51\\
53791 & 436 &   \textrm{Polarization 1} & 1.22 $\pm$0.03 & 7.2$\pm$0.2 & 26.5$\pm$0.3 & 35\\
53791 & 439.11 &  \textrm{Polarization 1}& 1.11$\pm$0.03 & 5.8$\pm$0.2 & 29.8$\pm$0.4 & 55\\
53847 & 428 &  \textrm{Polarization 1} & 1.20$\pm$0.02 & 7.8$\pm$0.1 & 24.4$\pm$0.2 & 76\\
53873 & 428 &  \textrm{Polarization 1} & 1.17$\pm$0.02 & 7.2$\pm$0.1 & 25.7$\pm$0.2 & 117 \\
53791 & 420.89 & \textrm{Polarization 2} &  1.19 $\pm$0.04 & 6.3$\pm$0.2 & 29.8 $\pm$0.5 & 35\\
53791 & 424 & \textrm{Polarization 2} & 1.19 $\pm$ 0.03 & 6.2$\pm$ 0.2 & 30.2$\pm$0.3 & 29\\
53791 & 428 & \textrm{Polarization 2} & 1.16 $\pm$ 0.02 & 6.1 $\pm$ 0.1 & 30.3$\pm$0.2 & 78\\
53791 & 432 & \textrm{Polarization 2} & 1.13 $\pm$0.02 & 6.3 $\pm$0.1 & 28.2 $\pm$0.3 & 51\\
53791 & 436 &  \textrm{Polarization 2} & 1.16 $\pm$0.03 & 7.1$\pm$0.2 & 25.5$\pm$0.3 & 35\\
53791 & 439.11 &  \textrm{Polarization 2} & 1.05$\pm$0.03 & 5.7$\pm$0.2 & 29.3$\pm$0.3 & 55\\
53847 & 428 & \textrm{Polarization 2} & 1.19$\pm$0.02 & 7.8$\pm$0.1 & 24.2$\pm$0.2 & 76\\
53873 & 428 & \textrm{Polarization 2} & 1.17$\pm$0.02 & 7.2$\pm$0.1 & 25.7$\pm$0.2 & 117 
\enddata

\tablecomments{Weighted means and standard deviations for $C_1$, scintillation bandwidth, $\Delta \nu_{\text{d}}$, and scattering delay, $\tau_{\text{d}}$, and number of unique measurements/subscans incorporated into those weighted averages, $N_{\text{meas}}$, for all epoch-observing frequency combinations in our study considered over both individual polarization and total intensity. Note that we used the weighted average of individual scattering delays and scintillation bandwidths in a given epoch-frequency combination for calculating our $C_1$ weighted means and standard deviations. Consequentially, but statistically insignificantly, results may vary from if one were to instead calculate $C_1$ using the weighted means and standard deviations listed in the above table for the corresponding scintillation bandwidth and scattering delay.}

\end{deluxetable*}
\par Furthermore, if we can make the assumption that individual $C_1$ measurements and scintillation noise are correlated within a given epoch but not between them due to the large number of scintles across each band, we can create a total weighted average $C_1$ for all of our measurements at 428 MHz using the $C_1$ values from the individual weighted averages. This gives us an aggregate weighted average $C_1$ at 428 MHz of 1.21$\pm$0.01 for the total intensity data, consistent with a thick screen geometry just over 10\% the line of sight distance to the pulsar with a square-law wavenumber spectrum, assuming the screen is located at the halfway point along the line of sight \citep[10\% corresponds to $C_1 = 1.13$;][]{Cordes_1998}. Additionally, our precision allows us to rule out a thin screen geometry with a square-law medium ($C_1 = 1.00$) to 21$\sigma$, a thin screen geometry with a Kolmogorov medium ($C_1 = 0.957$) to 23$\sigma$, a uniform geometry with a square-law medium ($C_1 = 1.53$) to 32$\sigma$, and a thick screen geometry greater than 30\% of the line of sight distance to the pulsar with a square-law medium ($C_1 = 1.56$) to 35$\sigma$. While additional systematics would certainly become important to consider at these significance levels, it is incredibly unlikely they would be present to a degree that reduces our claims  below 5$\sigma$ certainty, as they would need to be 4.1 times the statistical uncertainty (assuming these errors could be added in quadrature) to decrease our least constrained result to 5$\sigma$, and 6.3 times the statistical uncertainty for our most constrained result.

We note that significant S/N falloff toward the edges of the band in our highest frequency polyphase filterbank on MJD 53791 resulted in transfer functions with asymmetric intensities across the band, leading to poorer quality autocorrelation functions and fits in our 439 MHz data. This issue was further exacerbated by our decision to remove half of the band with the most significant S/N falloff, resulting in lower integrated S/N due to the reduced bandwidth. While the quality of the recovered pulse broadening functions remains high, we believe the autocorrelation function fits in the 439 MHz data bias scintillation bandwidths, and thus measured $C_1$ values at 439 MHz, lower than their true values.

\par The thick screen geometries we considered were determined under the assumption of a line of sight electron density wavenumber spectrum described by a square-law medium. However, previous studies have found that the line of sight towards this pulsar is generally consistent with a Kolmogorov medium \citep{cordes_90, Kaspi_94, Ramachandran_2006, Lam_2016}, which may support the existence of an inner scale. Indeed, the same $C_1$ values associated with our conclusions could also arise from a Kolomogorov electron density wavenumber spectrum with an inner scale larger than the diffraction scale at relevant observing frequencies \citep{Cordes_1998, geiger2024nanograv125yeardataset}. The latter scenario is true in the very strong scattering regime \citep{cl+91} and may be accompanied by a change in the spectral index. 
\par At the observing frequencies used for spectral index measurements in \cite{Kaspi_94} and \cite{Lam_2016}, one would expect the inner scale to be significantly smaller than the diffraction scale. However, unlike the inner scale, the diffraction scale is frequency-dependent, decreasing at lower frequencies and eventually becoming less than the inner scale at very low frequencies. 
\par By inferring $C_1$, we are able to calculate an observation's diffraction scale, $l_d$, given by 
\begin{equation}
    \label{diff_scale}
    l_d = \frac{1}{\nu}\Bigg[\frac{cD\Delta \nu_{\rm d}}{4\pi C_1}\Bigg]^{1/2}W_{D,\rm ISS},
\end{equation}
where $\nu$ is the observing frequency, $c$ is the speed of light, $D$ is the pulsar distance, and $W_{D,\rm ISS}$ is a weight factor of order unity \citep{Cordes_1998}. Our calculated diffraction scales, assuming $D=2.9\pm0.3$ kpc \citep{1937_dist} and $W_{D,\rm ISS}=1.4$ are shown in Table \ref{table_diff_scale}. Overall, these measurements suggest a diffraction scale of around 11$\times 10^3$ km at these observing frequencies, with an aggregate weighted average of 11.5$\pm$0.4$\times10^3$ km for our 428 MHz observations. If our data were in the very strong scattering regime \citep{cl+91}, then the inner scale would be greater than the diffraction scale and we could place lower limits. As it stands, we are likely in a frequency regime where the inner scale is comparable to, but still less than, the diffraction scale, suggesting an inner scale for this line of sight on the order of $10^3$ km. These findings are consistent with \cite{Kaspi_94}, who place an upper limit on the inner scale along this line of sight at 20$\times10^{3}$ km, as well as \cite{Ramachandran_2006}. This is also comparable to other lines of sight where an inner scale has been constrained; \cite{geiger2024nanograv125yeardataset} performed a $\chi^2$ analysis considering a variety of geometries, wavenumber spectra, and inner scale-diffraction scale ratios to determine the best pulse broadening function for the millisecond pulsar J1903+0327, and found the most likely value for this line of sight's inner scale to be around 1.4$\times10^3$ km.

\begin{deluxetable}{CCC}
\tablecolumns{6}

\tablecaption{Measured Diffraction Scales \label{table_diff_scale}}
\tablehead{ \colhead{MJD} &
\colhead{Frequency} & \colhead{$\overline{l_d}$}
\vspace{-.2cm}
 \\  \colhead{} & \colhead{\text{(MHz)}} & \colhead{($10^3$\text{km})}}
 \startdata
53791 & 420.89 & 11.1$\pm$0.6\\
53791 & 424 & 10.9$\pm$0.6\\
53791 & 428 & 10.8$\pm$0.6\\
53791 & 432 & 11.0$\pm$0.6\\
53791 & 436 & 11.5$\pm$0.6\\
53791 & 439.11 & 10.6$\pm$0.6\\
53847 & 428 & 12.1$\pm$0.6\\
53873 & 428 & 11.8$\pm$0.6
\enddata

\tablecomments{Weighted means and standard deviations for $l_d$ in total intensity data.}

\end{deluxetable}

\section{Conclusions \& Future Work}\label{conclusions}
We used cyclic spectroscopy to perform the first ever inference of the scintillation parameter $C_1$ without prior assumptions of the pulse broadening function shape when performing deconvolution. At 428 MHz we measured an aggregate weighted average $C_1$ of 1.21$\pm$0.01 along the line of sight towards PSR B1937+21, consistent with  a thick screen geometry comprising roughly 10\% the Earth-pulsar distance with either a square-law medium or a Kolmogorov medium with an inner scale. We also rule out various thin screen geometries, as well as thick screen geometries comprising more substantial portions of the line of sight, with greater than $5\sigma$ certainty.
\par This marks a major advancement in the ability of pulsar astronomers to study the structure of the ionized ISM. Future measurements of this parameter, in conjunction with scintillation arc or secondary wavefield analyses, may allow for more informed interpretations regarding the origins of scattering screens, which are currently a source of considerable speculation towards most lines of sight \citep{ocker,stock}. Additionally, constraints on $C_1$ could be used to directly infer scattering delays from measured scintillation bandwidths, assuming $C_1$ measurements are properly scaled with observing frequency. Such inclusions in PTA noise models would alleviate biases in existing scattering delay monitoring efforts caused by incorrect $C_1$ assumptions, and could significantly improve PTA sensitivity to nanohertz frequency gravitational wave signals in their data.
\par Demonstrations of cyclic spectroscopy's capabilities, such as the one in this study, are essential for increasing adoption, as this field is still in its infancy. Furthermore, utilization of new systems such as the Green Bank Telescope cyclic spectroscopy backend, which will be fully operational in the fall of 2026, and current exploration into similar systems for instruments like LOFAR2 and DSA, will significantly improve accessibility for this computational and storage-intensive technique.
\section{Acknowledgments} 
\par The National Radio Astronomy Observatory and Green Bank Observatory are facilities of the U.S. National Science Foundation operated under cooperative agreement by Associated Universities, Inc. Thanks to Paul Demorest for providing these data and for paper comments, as well as for helpful discussions, particularly regarding aliasing. Thanks to Michael Lam and Reynier Squillace for valuable discussion on weighted averages using data from independent populations. Thanks to Timothy Dolch, Ryan Lynch, and Dan Stinebring for providing paper feedback.

\par \textit{Software}: \textsc{dspsr} \citep{dspsr}, \textsc{pycyc}, \textsc{numpy} \citep{numpy}, and \textsc{matplotlib} \citep{matplotlib}.

\bibliography{turner_1937_cs.bib}{}
\bibliographystyle{aasjournal}
\end{document}